\documentclass[final,5p,times,twocolumn]{elsarticle}
\usepackage[utf8]{inputenc}
\usepackage[english]{babel}
\usepackage{cmap}
\usepackage{amsmath}
\usepackage{graphicx}
\usepackage{ amssymb }
\usepackage{hyperref}
\usepackage{csquotes}
\usepackage{caption}
\usepackage{subcaption}
\usepackage{aas_macros}
\hypersetup{pdfauthor=author}
\usepackage{xcolor}
\usepackage{comment}

\journal{Physics Letters B}

\usepackage[normalem]{ulem}

\makeatletter
\def\ps@pprintTitle{%
  \let\@oddhead\@empty
  \let\@evenhead\@empty
  \let\@oddfoot\@empty
  \let\@evenfoot\@oddfoot
}
\makeatother

\begin{document}

\begin{frontmatter}

\title{Photon-axion mixing in thermal emission of isolated neutron stars}

\author[MSU]{Aleksei Zhuravlev\corref{cor1}}
\ead{zhuravlev.aa18@physics.msu.ru}
\cortext[cor1]{Corresponding author.}

\affiliation[MSU]{organization={Faculty of Physics, Lomonosov Moscow State University},
            addressline={Vorobjevy Gory 1}, 
            city={Moscow},
            postcode={119234},
            country={Russia}}

\author[MSU,SAI]{Sergei Popov}

\affiliation[SAI]{organization={Sternberg Astronomical Institute, Lomonosov Moscow State University},
            addressline={Universitetskij pr. 13}, 
            city={Moscow},
            postcode={119234}, 
            country={Russia}}

\author[SAI,INR,PRAO]{Maxim Pshirkov}
\affiliation[INR]{organization={Institute for Nuclear Research of Russian Academy of Sciences},
            addressline={60th October Anniversary Prospect, 7a}, 
            city={Moscow},
            postcode={117312}, 
            country={Russia}}
            
\affiliation[PRAO]{organization={P. N. Lebedev Physical Institute of the Russian Academy of Sciences Pushchino Radio Astronomy Observatory},
            addressline={PRAO}, 
            city={Pushchino},
            postcode={142290}, 
            country={Russia}}

\begin{abstract}

Thermally emitting neutron stars represent a promising environment for probing the properties of axion-like particles. Due to the strong magnetic fields of these sources, surface photons may partially convert into such particles in the large magnetospheric region surrounding the stars, which will result in distinctive signatures in their spectra. However, the interaction depends on the polarization state of the radiation and is rather weak due to the low experimentally allowed values of the coupling constant $g_{\gamma a}$. In this work, we compute the degree of photon-axion transition in the case of 100\% O-mode polarization and spectral energy distribution of an isotropic blackbody with uniform surface temperature. The stellar magnetic field is assumed to be dipolar. We show that with the maximum effect reached for the magnetic fields $\sim10^{13}$~--~$10^{14}$~G (typical for X-ray dim isolated neutron stars) and $g_{\gamma a} = 2 \times 10^{-11}$ GeV$^{-1}$, the optical flux is reduced by 30~--~40\%, while the high-energy part of the spectrum is not affected. The low-energy decrease exceeds 5\% at $g_{\gamma a} \geq 2 \times 10^{-12}$ GeV$^{-1}$ and $m_a \leq 2\times 10^{-6}$ eV, which is below the present experimental and astrophysical limits on axion parameters. To obtain the actual observational constraints, rigorous treatment of the radiative surface layers is required.

\end{abstract}


\end{frontmatter}

\section{Introduction}

Pseudoscalar axion-like particles appear in various extensions of the Standard Model of particle physics. Proposed in the late 1970s to explain the absence of CP violation in strong interactions \cite{1977PhRvL..38.1440P, 1978PhRvL..40..279W}, they have transcended the original context and were employed to solve a variety of problems in physics and astrophysics \cite{2020PhR...870....1D,2020arXiv201205029C,2016PhR...643....1M,raffelt1996stars}. In this work, we refer to all such particles simply as axions.

The defining property of axions is the conversion into photons and vice versa in the presence of an external magnetic field \cite{1988PhRvD..37.1237R}. Many laboratory-based searches exploited this interaction in attempts to prove their existence, but so far no satisfactory observational evidence has been found, and the stringent bounds on the coupling constant over a wide axion mass range were obtained by the CERN Axion Solar Telescope (CAST), $g_{\gamma a} \leq 6.6 \times 10^{-11}$ GeV$^{-1}$ (95\% C.L.) \cite{2017NatPh..13..584A}. For this reason, studying manifestations of axions in various astrophysical environments has become increasingly popular. A common strategy used in such works is to search for spectral features in sources of radiation observed through large-scale regions with relatively strong magnetic fields. These include active galactic nuclei such as NGC 1275 \cite{2016PhRvL.116p1101A,2017ApJ...847..101B,2018MNRAS.479.2243C} (see also \cite{2020PhLB..80235252L}), blazars and quasars \cite{2011PhRvD..84l5019F,2019MNRAS.487..123G}, supernovae \cite{Payez:2014xsa}, etc. By comparing the detected spectra with the ones arising after the mixing was taken into account, several authors were able to impose astrophysical constraints on a range of axion parameters, exceeding the CAST limit.

More recently, considerable research has been devoted to studying the axion-photon interaction in strongly magnetized objects, namely magnetic white dwarfs (mWDs) and neutron stars. In the case of white dwarfs, photons emitted by the photosphere may undergo conversion as they propagate through the encompassing gaseous atmosphere \cite{2011PhRvD..84h5001G}. If the plasma is sufficiently tenuous, this could lead to an enhancement of the observed degree of linear polarization. Since in most white dwarfs its value is very low, such interaction can be used to constrain a range of axion parameters. In particular, for mWD PG 1015+014 the induced polarization degree should exceed the observed value of 5\% at $g_{\gamma a} \geq 2 \times 10^{-11}$ GeV$^{-1}$, which is below the CAST limit.

In the case of neutron stars, the presence of axions can manifest itself in a number of ways. First, their magnetospheres can be a site for the conversion of dark matter axions \cite{2009JETP..108..384P, 2020PhRvL.125q1301F}. Also, axions could be produced thermally inside the core with energies from several keV to several hundred keV through the nucleon bremsstrahlung processes, and escape the interior due to the feeble interaction with matter. Their subsequent transition into photons in the magnetic field surrounding the star could lead to observable signatures in its hard X-ray spectrum \cite{raffelt1996stars,RAFFELT19901,2019PhRvD..99d3011S}. Magnetars are natural targets for such investigation due to the extremely strong magnetic fields and high internal temperatures, and their study led to further astrophysical limits on axion parameters \cite{2018JHEP...06..048F,2019JHEP...01..163F,2021PhRvD.103b3010L}.

At the same time, manifestations of the reverse process have not yet been fully explored. Spectra of neutron stars commonly include a thermal component, and as the surface radiation propagates through the large magnetospheric region with a relatively strong magnetic field, a fraction of the initial flux may oscillate into axions. Although in most of the magnetosphere the probability of conversion is suppressed by the QED effects, previous studies have shown that the mixing can lead to distinctive spectral signatures \cite{2006PhRvD..74l3003L,2012ApJ...748..116P}. However, the coupling values they considered are now excluded experimentally. The purpose of our work is to find out whether current constraints on axion parameters allow for observable effects, and to estimate their possible magnitude.

X–ray dim isolated neutron stars (XDINSs), also known as the “Magnificent Seven”, are particularly suited for such investigation. These isolated, radio-silent sources exhibit purely thermal spectra with no high-energy power-law component, which is often found in other classes of neutron stars. Their X–ray luminosity of $10^{31} - 10^{32}$ erg s$^{-1}$ is consistent with the surface blackbody emission with temperatures $kT^{\infty} \sim 40 - 100$ eV, and the radiation radii of a few kilometers \cite{2009ApJ...705..798K,2009ASSL..357..141T}. The spin periods $P \sim 3 - 12$ s and period derivatives $\dot{P} \sim 10^{-14} - 10^{-13}$ s s$^{-1}$ translate into spin-down magnetic fields $B \sim 10^{13} - 10^{14}$ G.

In the strongly magnetized vacuum around a neutron star, the photons propagate in two normal modes, the extraordinary (X) and ordinary (O) mode, and only those polarized in the latter can convert into axions. Although the physical conditions of the outermost layers of XDINSs are still a debated issue, several models have been proposed to explain the observed properties of these sources  \cite{2014PhyU...57..735P}, and the polarization patterns they predict are different. The conventional picture is that cooling neutron stars are covered with a gaseous atmosphere that reprocesses the emission coming from the outermost surface layers. As a result, the radiation acquires a net polarization predominantly in the X-mode, since the strong magnetic field substantially reduces its opacity while leaving that for the O-mode almost unchanged \cite{2016MNRAS.459.3585G,2006RPPh...69.2631H, 2006MNRAS.373.1495V}. On the other hand, it has recently been appreciated that the stars with a low surface temperature ($\lesssim  100 $ eV) and a strong magnetic field ($\gtrsim 10^{13}$ G) may have a liquid or solid condensed surface due to a phase transition in the outermost layers \cite{2009ASSL..357..141T,1997ApJ...491..270L,2001RvMP...73..629L,2007MNRAS.382.1833M}. Thus, in this case, the atmosphere may be absent. The emissivity of a naked condensed surface in both X- and O-mode is of the same order in a broad energy range \cite{2016MNRAS.459.3585G,2010A&A...522A.111S}.

Since the magnitude of photon-axion transition will depend on the composition of stellar surface layers, it is essential to estimate its possible magnitude at the currently allowed values of axion parameters. To this end, here we consider a simple picture in which the thermal emission is an isotropic blackbody with a uniform surface temperature, and is 100\% polarized in the O-mode. These conditions provide the highest degree of conversion, and our results will show whether a more detailed study can lead to astrophysical constraints that exceed the present experimental limits. 

Our paper is organized as follows. In Section 2, we summarize the main properties of the photon-axion interaction in a strong magnetic field. Then in Section 3, we calculate the resulting degree of conversion and determine the range of axion parameters $ \left (m_a, g_{\gamma a} \right) $ at which the mixing has a significant effect. The observability of our results and the prospects for future work are discussed in Section 4.

\section{Theoretical framework}

In this section we outline the physical basis for our calculations. Throughout the paper we use Lorentz–Heaviside units with $\hbar = c = 1$ and $\alpha = e^2/4\pi \approx 1/137$.

\subsection{Polarization of radiation}

In the presence of an external magnetic field, the vacuum around the star behaves as a birefringent medium, in which the photons are polarized in two normal modes: the ordinary mode (O-mode), with the electric field oscillating in the plane of the propagation vector $\boldsymbol{k}$ and the local magnetic field $\boldsymbol{B}$, and the extraordinary mode (X-mode), with the electric field oscillating perpendicularly to both $\boldsymbol{k}$ and $\boldsymbol{B}$. According to the QED, the dielectric and magnetic permeability tensors of the vacuum are modified by virtual electron-positron pairs, which affects the polarization properties of radiation primarily in two ways. 

First, the photons emitted at the surface maintain their initial polarization state up to a characteristic distance \cite{2002PhRvD..66b3002H,2003MNRAS.342..134H,2015MNRAS.454.3254T}, the polarization radius $r_p$, which depends on the photon energy and magnetic field strength. In this region, the wave electric field can instantly adapt its direction to that of $\boldsymbol{B}$. Around $r_p$, the coupling weakens, and the photon electric field is no longer able to properly follow the variation of the local magnetic field, until for $r \gg r_p$, its direction freezes. If the radiation propagates radially in a dipolar field, its initial polarization state is preserved because the direction of transverse part of $\boldsymbol{B}$ is always constant. This does not hold for the photons, trajectory of which has been modified by the gravitational effects \cite{2002ApJ...566L..85B}. In general, to account for the QED-induced depolarization at $r > r_p$, it is necessary to solve the system determining the evolution of the polarization state \cite{2015MNRAS.454.3254T, 2014MNRAS.438.1686T}. However, the magnitude of this effect is relatively small, because the dipolar magnetic field does not significantly change its direction outside the polarization radius. According to our calculations, for the energies considered in this paper, the angle between the wave electric field and the transverse part of $\boldsymbol{B}$ for both modes changes only slightly from its initial value in the region that will be of interest to us. For this reason, we can assume that the radiation always retains its initial mode without losing significant physical accuracy.

Second, the refractive indices for each mode are different and can deviate from their vacuum value of unity. For the photon energies well below the electron rest mass, the Euler-Heisenberg effective Lagrangian can be applied \cite{1971AnPhy..67..599A, 2004ApJ...612.1034P}, which for arbitrary values of $B = |\boldsymbol{B}|$ leads to
 
\begin{equation}
    \begin{Bmatrix} 
    n_\parallel \\ n_\perp
    \end{Bmatrix} = 1 +
    \sin^2\Theta 
    \begin{Bmatrix} 
    \xi \left( b \right) \\ \chi \left( b \right)
    \end{Bmatrix},
\end{equation}

\begin{equation}
    \xi\left( b \right) = \frac{7\alpha b^2 \left( 1 + 1.2 b \right)}{45\pi \left[ 1 + 1.33 b + 0.56 b^2 \right]},
\end{equation}
   
\begin{equation}
    \chi\left( b \right) = \frac{4\alpha b^2}{45\pi \left[ 1 + 0.72 b^{5/4} + \left( 4/15 \right) b^2 \right]},
\end{equation}

\noindent where $\Theta$ is the angle between $\boldsymbol{k}$ and $\boldsymbol{B}$, $b$ is the magnetic
field strength in units of the critical field $B_{\text{crit}} = m^2_e/e = 4.4\times10^{13}$ G.

The magnetized plasma surrounding the star modifies the dielectric tensor similarly to the vacuum polarization. In the following, we assume that it has the Goldreich-Julian charge density $\rho_{\text{GJ}} = -\boldsymbol{\Omega} \cdot \boldsymbol{B}$, where $\boldsymbol{\Omega}$ is the stellar angular velocity \cite{1969ApJ...157..869G}. Note that $\rho_{\text{GJ}}$ estimates only the net charge density, and the total charge density may be larger due to the presence of electron-positron pairs \cite{2006MNRAS.368.1377S}. However, since there is no reliable probe of the plasma density in the magnetospheres of neutron stars, we will only consider the Goldreich-Julian value as a conservative lower limit.

\subsection{Photon-axion mixing}

The oscillations of a photon with energy $\omega$ and an axion in the external magnetic field are described by the following interaction: 

\begin{equation}
    \mathcal{L} = - \frac{1}{4} g_{\gamma a} a F_{\mu \nu} \tilde{F}^{\mu \nu}, 
\end{equation}

\noindent where $F_{\mu \nu}$ is the electromagnetic stress tensor, $\tilde{F}^{\mu \nu}$ is its dual, $a$ is the axion field, $m_a$ is the axion mass, and $g_{\gamma a}$ is the coupling constant \cite{1988PhRvD..37.1237R}. By adding the above term to the Euler-Heisenberg effective Lagrangian and linearizing the wave equations, we obtain the following system:

\begin{align}\label{eqn_2_matrix_form}
    \left(\omega + \begin{pmatrix}
    \Delta_\parallel + \Delta_p & \Delta_{M}\\
    \Delta_{M}    & \Delta_{a}
    \end{pmatrix} - i\partial_z \right)
    \begin{pmatrix}
    A_\parallel \\ a
    \end{pmatrix} = 0,
\end{align}

\noindent where $A_\parallel$ and $a$ denote the amplitudes of the ordinary photon state and the axion, respectively, and the $z-$ axis is along $\boldsymbol{k}$. Note that only the radiation polarized in the O-mode can convert into axions, and the X-mode flux always remains intact. The components of the mixing matrix are:

\begin{equation}
\begin{aligned}
\Delta_{a} &= -\frac{m_a^2}{2\omega}, & \Delta_{M} &= \frac{1}{2} g_{\gamma a} B \sin\Theta, \\
\Delta_{p} &= -\frac{\omega_p^2}{2\omega}, & \Delta_{\parallel} &= \frac{1}{2} \omega\left( n_{\parallel} - 1 \right),
\end{aligned}
\end{equation}

\noindent where $\omega_{p}^{2}=4 \pi \alpha n_{e} / m_{e}$ is the plasma frequency, $n_{e} =  \rho_{\text{GJ}}/e$ is the electron density \cite{2007PhRvD..76l3011H}.

While the converted fraction of the initial photon flux can only be obtained by numerically integrating Equations \ref{eqn_2_matrix_form}, the strength of the mixing can be estimated as follows. The external magnetic field varies along the photon trajectory over a lengthscale $l_B = B/|\hat{\boldsymbol{k}}\cdot\nabla B| = r/3$, where a hat denotes a unit vector. If we take the value of $B$ to be constant over $l_B$, the probability of a photon-axion transition after traveling this distance is

\begin{equation}\label{eqn_2_probability}
    P_{\gamma_{\parallel} \rightarrow a} = \frac{4\Delta_{M}^2}{\left(\Delta_\parallel + \Delta_{p} - \Delta_a\right)^2 + 4\Delta_M^2} \sin^2\left(\frac{1}{2}\Delta_{\text{osc}} l_B\right)
\end{equation}
    
\noindent where $\Delta_{\text{osc}}^2 = \left(\Delta_\parallel + \Delta_{p} - \Delta_a\right)^2 + 4\Delta_{M  }^2$, so that the oscillation length is $l_{\text{osc}} = 2\pi/\Delta_{\text{osc}}$. Strong mixing occurs when the conversion probability is close to unity, which is satisfied if

\begin{equation}\label{eqn_3_conditions}
    \begin{aligned}
        l_{\text{osc}} &\lesssim l_B,  \\
        |\Delta_\parallel + \Delta_{p} - \Delta_a| &\lesssim 2 |\Delta_{M}|.
    \end{aligned}
\end{equation}

\section{Numerical simulations}

In this section, we calculate the observational signatures of photon-axion mixing in the O-mode flux of a thermally emitting neutron star, and determine the range of axion parameters $ \left (m_a, g_{\gamma a} \right) $ at which the effect is significant.

\subsection{Computational model}

Our treatment of the surface radiation relies on a number of simplifications which are discussed below. We stress that our primary goal is to assess whether the mixing can have a noticeable effect on spectra of XDINSs at the currently unconstrained values of $ \left (m_a, g_{\gamma a} \right) $ and whether there is room for future investigations, rather than to derive theoretical predictions to be compared with the observational data. For this reason, we avoid dealing with the complex structure of the outermost stellar layers and consider the most favorable conditions for the detection of photon-axion conversion. 

The stellar magnetic field is assumed to be an aligned dipole, components of which in polar coordinates $\left( r, \theta \right)$ take the form

\begin{equation}
    \boldsymbol{B}^{\text{dip}} = \frac{B_p}{2} \left(\frac{R_s}{r}\right)^3 \begin{pmatrix}
    2 f_{\text{dip}} \cos\theta \\
    g_{\text{dip}} \sin\theta \\
    0
    \end{pmatrix},
\end{equation}

\noindent where $B_p$ is the surface polar field strength. The functions $f_{\text{dip}}$, $g_{\text{dip}}$ account for the relativistic corrections \cite{1986SvA....30..567M, 1996ApJ...473.1067P}. The stellar radius, mass, and period are $R_s = 12$ km, $M_s=1.4 M_\odot$, and $P_s = 7$ s, respectively.

Initially, the radiation is emitted from the cooling stellar surface with an isotropic blackbody distribution and uniform temperature $k T^\infty = 50$ eV. The seed photons are $100\%$ polarized in the O-mode, which provides the maximum degree of conversion into axions. As the radiation moves away from the emission point, we numerically integrate Equations \ref{eqn_2_matrix_form} along its trajectory, accounting for the gravitational redshift. The observer is located at infinity, with the line of sight inclined at an angle $\chi$ with respect to the spin axis, and collects the contribution of each part of the surface which is into view due to the effects of ray bending \cite{2002ApJ...566L..85B}.

\subsection{Results}

\begin{figure}
    \centering
    \includegraphics[width=\linewidth]{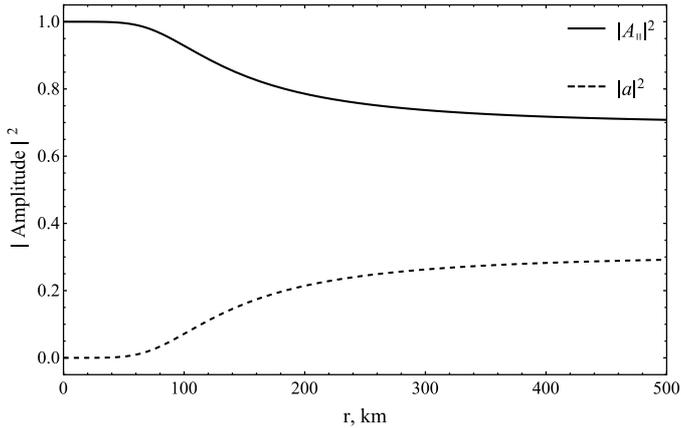}
    \caption{Evolution of the photon and axion square amplitudes in the case of radiation emitted radially, as a function of the distance $r$ from the stellar surface. The parameters of the system are $\omega_\infty = 1$ eV, $g_{\gamma a} = 2\times 10^{-11}$ GeV$^{-1}$, $m_a = 10^{-8}$ eV. The magnetic field geometry is an aligned dipole with $B_p = 10^{13}$ G and $\chi = 60^\circ$. The probability of conversion beyond $r = 500$ km is vanishingly small, this region is not displayed.}
    \label{fig_1}
\end{figure}

To get a clear picture of the conversion process, let us first study the mixing in an individual photon ray. The simplest case is the radiation emitted in radial direction (colatitude of the emission point is the same as that of the observer), so that its trajectory is not modified by gravity. We take the coupling constant $g_{\gamma a} = 2\times 10^{-11}$ GeV$^{-1}$ (the highest value allowed by the study of mWD PG 1015+014 \cite{2011PhRvD..84h5001G}), axion mass $m_a = 10^{-8}$ eV, and photon energy $\omega_\infty = 1$ eV. The polar magnetic field strength is $B_p = 10^{13}$ G (representative of XDINSs \cite{2009ASSL..357..141T}) and the observer inclination angle is $\chi= 60^\circ$. The resulting plot of the photon and axion square amplitudes is shown in Figure \ref{fig_1}.

As we can see, significant conversion takes place at a distance of several stellar radii away from the surface. The system does not enter the strong-oscillation regime (in this case, $|A_\parallel|^2$ and $|a|^2$ would have the form of sinusoidal curves) because Conditions \ref{eqn_3_conditions} are never met simultaneously. In particular, near the emission point, where $l_{\text{osc}}$ is small compared to $l_B$, $\Delta_\parallel$ is much larger than $2 \Delta_M$, and therefore the mixing is suppressed. As the radiation moves away from the surface of the star, $\Delta_\parallel$ eventually becomes comparable with $2 \Delta_M$ ($\Delta_a$ is negligible and $|\Delta_{p}| \lesssim \Delta_\parallel$), which is due to their different dependence on the radial coordinate: $\Delta_\parallel \sim 1/r^6$, while $\Delta_M \sim 1/r^3$. However, in this region, $l_{\text{osc}}$ is already 5-10 times greater than $l_B$, which results in the conversion probability being less than unity. Still, these conditions are sufficient for the mixing to have a noticeable effect. At relatively large distances, the conversion probability becomes negligible.

\begin{figure}
    \centering
    \includegraphics[width=\linewidth]{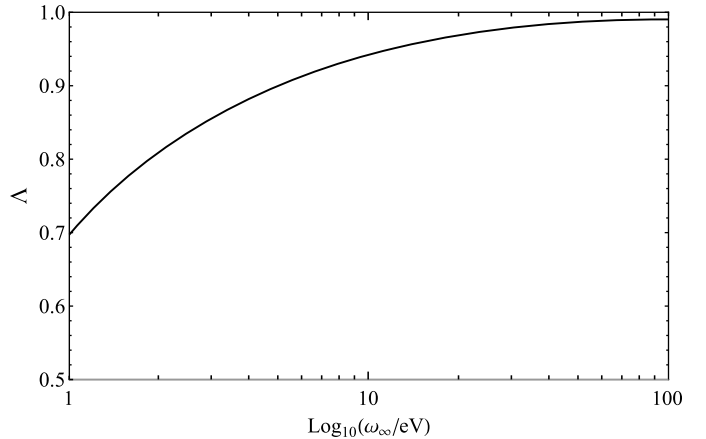}
    \caption{Differential modification factor $\Lambda$ of the O-mode radiation for a broad energy range. The remaining parameters are the same as in Figure \ref{fig_1}. At $\omega_\infty > 100$ eV, $\Lambda$ is almost unity, this region is not displayed.}
    \label{fig_2}
\end{figure}

We then perform the calculations for a broad energy range (from 1 eV to 10 keV), integrating the intensity over the visible part of the stellar surface. The result is conveniently expressed in terms of the differential modification factor $\Lambda$, which is defined as the fraction of the initial photon flux with a given energy, reaching the observer without conversion. In particular, if $\Lambda = 1$, the flux will not be modified, and if $\Lambda = 0$, all photons will turn into axions and thus no signal will be detected. The data for $g_{\gamma a} = 2\times 10^{-11}$ GeV$^{-1}$, $m_a = 10^{-8}$ eV, and $B_p = 10^{13}$ G, $\chi = 60^\circ$ are presented in Figure \ref{fig_2}, as a function of the redshifted photon energy. 

Our results show that the mixing is most significant at the optical energies, where the O-mode flux is almost halved. At the same time, it has little effect on the X-ray emission, which is due to the fact that the QED contribution $\Delta_\parallel$ increases linearly with increasing $\omega$, while $\Delta_M$ does not depend on it. At higher photon energies, $\Delta_\parallel$ and $\Delta_M$ become of the same order in the region where the oscillation length is much larger than $l_B$, which results in a considerable decrease in the conversion probability. The shapes of the photon and axion square intensities are similar to those shown in Figure \ref{fig_1}, except for the oscillations having smaller amplitudes. It is noteworthy that the converted fraction of the flux varies by no more than a few percent between different ray trajectories, which due to the fact that the system remains in the weak-oscillation regime. For the same reason, its value does not change noticeably for other observer inclination angles.

\begin{figure}
  \centering
    \includegraphics[width=\linewidth]{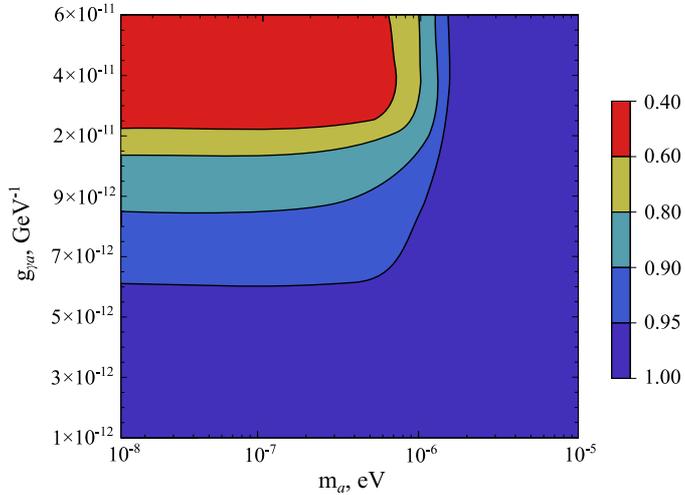}
    \caption{Contour plot of the O-mode modification factor $\Lambda$ for $\omega_\infty = 1$ eV, as a function of the axion parameters $\left( m_a, g_{\gamma a} \right)$. The magnetic field geometry is $B_p = 10^{13}$ G, $\chi = 60^\circ$. We also include the values of $g_{\gamma a}$ allowed by CAST \cite{2017NatPh..13..584A}.}
    \label{fig_3}
\end{figure}

We then repeat our calculations for different values of $\left( m_a, g_{\gamma a} \right)$ to obtain the range at which the mixing has a significant effect on the optical radiation. Figure \ref{fig_3} illustrates the modification factor at $\omega_\infty = 1$ eV for $B_p = 10^{13}$ G and a given choice of axion parameters. The optical photon flux is reduced by more than 5\% at $g_{\gamma a} \geq 6 \times 10^{-12}$ GeV$^{-1}$ and $m_a < 2\times 10^{-6}$ eV. This result can be explained using Conditions \ref{eqn_3_conditions}: for lower values of the coupling constant, $\Delta_{M}$ does not get close enough to $\Delta_\parallel$ in the region with a relatively small oscillation length; for higher axion masses, $\Delta_a$ becomes of the same order as both $\Delta_{M}$ and $\Delta_\parallel$, suppressing the conversion probability. The plasma contribution $\Delta_{p}$ becomes important at lower values of $g_{\gamma a}$, where $\Delta_\parallel \gg \Delta_{M}$ almost everywhere in the magnetosphere. At some point, the terms $\Delta_p$ and $\Delta_\parallel$ become of the same order and cancel each other because of their opposite signs. Even though the oscillation length in this region is much larger than $l_B$, this results in a slight increase in the conversion probability.  

\begin{figure}
  \centering
    \includegraphics[width=\linewidth]{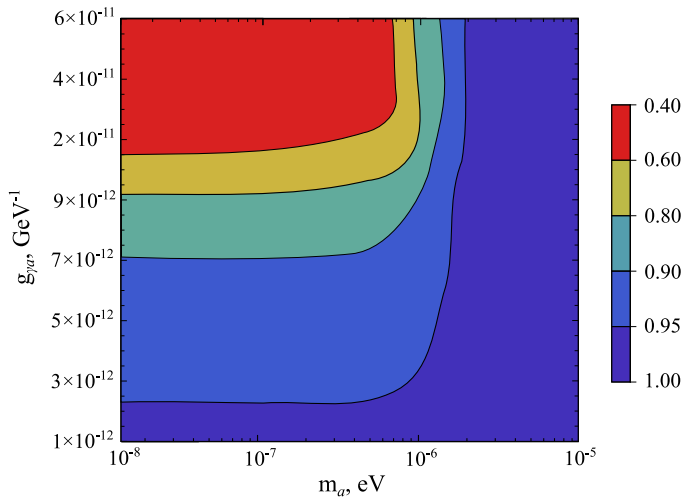}
    \caption{Same as Figure \ref{fig_3}, but for $B_p = 10^{14}$ G.}
    \label{fig_4}
\end{figure}

We also perform the same calculations for a greater polar field strength $B_p = 10^{14}$ G, which can also be present in XDINSs \cite{2009ASSL..357..141T}. As shown in Figure \ref{fig_4}, in this case the converted fraction reaches 5\% at even lower values of the coupling constant, $g_{\gamma a} \geq 2 \times 10^{-12}$ GeV$^{-1}$ and $m_a \leq 2\times 10^{-6}$ eV.

\section{Discussion and conclusions}

In this paper we studied the photon-axion conversion in the magnetosphere of a thermally emitting neutron star. Assuming the dipolar magnetic field and the spectral energy distribution of an isotropic blackbody with a uniform surface temperature, we have shown that the mixing can result in a significant decrease in the optical O-mode flux at the allowed values of $\left( m_a, g_{\gamma a} \right)$. Further investigation may lead to astrophysical constraints on axion parameters that exceed the present experimental and astrophysical limits.

We found that the photon-axion system exhibits only weak oscillations at the currently allowed values of the coupling constant. The reason for the decrease in the conversion probability is either the dominance of the QED contribution $\Delta_\parallel$, or the fact that the oscillation length is much larger than the scale length of the magnetic field. This is in agreement with the study of \cite{2012ApJ...748..116P}, who focused on higher photon energies and currently excluded values of $g_{\gamma a}$. According to their results, weak oscillations are present in the X-ray range at $g_{\gamma a} = 10^{-10} - 10^{-9}$ GeV$^{-1}$ and the converted fraction depends inversely on $\omega_\infty$, similarly to the case shown in Figure \ref{fig_2}. Only at a relatively high value of $g_{\gamma a} = 10^{-8}$ GeV$^{-1}$, the conversion probability becomes equal to unity in a large area of the magnetosphere, and the system enters the strong-oscillation mode. As a result, $\Lambda$ depends on the photon energy in a more complex way and, in particular, can increase with increasing $\omega_\infty$. Such relation should not be confusing, since it is simply a consequence of the strong-oscillation mode, which is substantially different from the weak-mixing regime discussed in our work \cite{1988PhRvD..37.1237R,2006PhRvD..74l3003L}. Overall, greater values of the coupling constant could be more easily constrained by looking into the X-ray energies where the emission of XDINSs typically peaks, while at lower values of $g_{\gamma a}$ one has to consider the mixing in the optical band.

Our results exceed the limits obtained by \cite{2011PhRvD..84h5001G} in their study of mWD PG 1015+014. At the same time, observations of the magnetic white dwarfs with higher surface field strengths could potentially exclude a wider region of the axion parameter space, which will be surpassed by the limits shown in Figures \ref{fig_3} and \ref{fig_4} only at high axion masses. However, these results should be interpreted with caution, as they are highly dependent on the assumptions about the atmospheric plasma composition. In particular, the authors of \cite{2011PhRvD..84h5001G} consider the case of a relatively sparse hydrogen plasma with a surface density of $\rho_0 = 10^{-10}$ g cm$^{-3}$ and a barometric profile. They also note that there is no clear agreement on the surface plasma density of magnetic white dwarfs, and its actual value is expected to lie between $10^{-11}$ and $10^{-6}$ g cm$^{-3}$. In this regard, the study of XDINSs may provide a more robust method of imposing constraints on axion parameters, which, however, will depend on another kind of assumptions.

A major challenge in assessing the effect of mixing on spectra of XDINSs is the uncertainty in the structure of their radiative surface layers. A well-known feature of the "Magnificent seven" is that their optical and ultraviolet counterparts exceed the extrapolation of X–ray blackbody at low energies by a factor 5 – 50 \cite{2011ApJ...736..117K}, and numerous authors have made extensive use of  atmospheric and condensed surface models in attempts to explain its origin. A strongly magnetized atmosphere has been suggested for RX J1605.3+3249 \cite{2007MNRAS.377..905M}. The observed properties of RX J0720.4-3125 were attributed to the emission of a bare condensed surface \cite{2006A&A...459..175P}, and a combination of the latter and a thin, magnetic, partially ionized hydrogen atmosphere was used to fit the spectral shape of J1856.5-3754 \cite{2007MNRAS.375..821H} and RX J1308.6+2127 \cite{2011A&A...534A..74H}. 

The polarization pattern induced by both models can be conveniently expressed in terms of the intrinsic polarization degree (that at the source): $\Pi_\text{L}^\text{EM} = \left( F_\text{X} - F_\text{O} \right) / \left( F_\text{X} + F_\text{O} \right)$, where $F_\text{X,O}$ is the monochromatic, phase-averaged flux in each mode, obtained by integrating the intensity over the visible part of the stellar surface \cite{2016MNRAS.459.3585G}. By definition, $\Pi_\text{L}^\text{EM} = 1$ for the radiation 100\% initially polarized in the X-mode, $\Pi_\text{L}^\text{EM} = -1$ for that in the O-mode, and $\Pi_\text{L}^\text{EM} = 0$ for the unpolarized emission. According to the results of \cite{2016MNRAS.459.3585G}, in the case of a fully ionized hydrogen atmosphere, $\Pi_\text{L}^\text{EM} \approx 0.99$ in the X–ray band and $\Pi_\text{L}^\text{EM} \approx 0.87$ in the optical band for all possible viewing geometries. The radiation of a condensed surface in the X-ray band is almost unpolarized, $|\Pi_\text{L}^\text{EM}| \lesssim 0.07$, while in the optical band $\Pi_\text{L}^\text{EM}$ can reach $-0.5$ for the most favorable viewing geometries, with the emission being mostly polarized in the O-mode. Since the polarization state of thermal radiation directly affects the magnitude of photon-axion conversion, the effect will be observable only if the flux contains a considerable fraction of O-mode photons. From this point of view, the emission of a condensed surface is a promising place to search for its manifestations, but the result is likely to be lower than that obtained in Section 3, since we considered the liming case of 100\% O-mode polarization. The presence of an atmosphere may make the interaction less noticeable because its radiation is predominantly polarized in the X-mode. 

Rigorous treatment of the outermost stellar layers provides several ways of imposing constraints on a range of axion parameters. One approach is to simply measure the change in the low-energy spectra, as shown in Figure \ref{fig_3}. Although the optical counterparts of XDINSs are rather faint, observational data have been presented for all sources \cite{2011ApJ...736..117K}. However, this method will not be sensitive enough if the converted fraction is relatively small, which is expected at low values of the coupling constant $g_{\gamma a}$. In such cases, it may be more efficient to compute the polarization observables, linear polarization fraction $\Pi_\text{L}$ and polarization angle $\chi_\text{p}$ \cite{2015MNRAS.454.3254T}, which accurately describe the polarization state of the detected radiation and can reflect a slight decrease in the O-mode flux. Since various types of stellar surface composition are expected to have different polarization patterns \cite{2016MNRAS.459.3585G}, constraints can be based on a comparison of $\Pi_\text{L}$ and $\chi_\text{p}$, predicted by the models which do and do not take the mixing into account. Additionally, one can make use of the observational polarimetric data, as described in \cite{2017MNRAS.465..492M} for the case of RX J1856.5-3754. 

Another potentially promising approach is to consider the photon-axion interaction in magnetars. Their magnetic fields are stronger than those of XDINSs $\left( B_p \approx 10^{15} \text{ G}\right)$, and are believed to include higher multipoles and global or local shears (the ``twisted magnetosphere'' model, see \cite{2002ApJ...574..332T}), which in general can increase the mixing strength. In addition, their low-energy radiation may include a considerable fraction of the O-mode photons \cite{Beloborodov_2007, 2020ApJ...904..184T}. Such investigation, however, is hampered by our present poor knowledge of both the composition of surface layers and the magnetospheric structure of magnetars, as well as the faintness of the optical counterparts of their thermal emission \cite{2015RPPh...78k6901T}. At the same time, considering other classes of neutron stars, such as radio pulsars, is unlikely to produce greater constraints on axion parameters. Due to their lower magnetic field strengths, the converted fraction will likely be smaller than in the case of XDINSs.

\section*{Acknowledgements}

We are grateful to Sergey Troitsky for numerous helpful discussions and the anonymous referee for insightful comments and suggestions. This work was supported by Russian Science Foundation, grant 18-12-00258.

\bibliographystyle{elsarticle-num} 
\bibliography{axions_rxj}

\begin{thebibliography}{10}
\expandafter\ifx\csname url\endcsname\relax
  \def\url#1{\texttt{#1}}\fi
\expandafter\ifx\csname urlprefix\endcsname\relax\def\urlprefix{URL }\fi
\expandafter\ifx\csname href\endcsname\relax
  \def\href#1#2{#2} \def\path#1{#1}\fi

\bibitem{1977PhRvL..38.1440P}
R.~D. {Peccei}, H.~R. {Quinn}, {CP conservation in the presence of
  pseudoparticles}, \prl 38~(25) (1977) 1440--1443.
\newblock \href {https://doi.org/10.1103/PhysRevLett.38.1440}
  {\path{doi:10.1103/PhysRevLett.38.1440}}.

\bibitem{1978PhRvL..40..279W}
F.~{Wilczek}, {Problem of strong P and T invariance in the presence of
  instantons}, \prl 40~(5) (1978) 279--282.
\newblock \href {https://doi.org/10.1103/PhysRevLett.40.279}
  {\path{doi:10.1103/PhysRevLett.40.279}}.

\bibitem{2020PhR...870....1D}
L.~{Di Luzio}, M.~{Giannotti}, E.~{Nardi}, L.~{Visinelli}, {The landscape of
  QCD axion models}, \physrep 870 (2020) 1--117.
\newblock \href {http://arxiv.org/abs/2003.01100} {\path{arXiv:2003.01100}},
  \href {https://doi.org/10.1016/j.physrep.2020.06.002}
  {\path{doi:10.1016/j.physrep.2020.06.002}}.

\bibitem{2020arXiv201205029C}
K.~{Choi}, S.~H. {Im}, C.~S. {Shin}, {Recent progresses in physics of axions or
  axion-like particles}, arXiv e-prints (2020) arXiv:2012.05029\href
  {http://arxiv.org/abs/2012.05029} {\path{arXiv:2012.05029}}.

\bibitem{2016PhR...643....1M}
D.~J.~E. {Marsh}, {Axion cosmology}, \physrep 643 (2016) 1--79.
\newblock \href {http://arxiv.org/abs/1510.07633} {\path{arXiv:1510.07633}},
  \href {https://doi.org/10.1016/j.physrep.2016.06.005}
  {\path{doi:10.1016/j.physrep.2016.06.005}}.

\bibitem{raffelt1996stars}
G.~G. Raffelt, Stars as laboratories for fundamental physics: The astrophysics
  of neutrinos, axions, and other weakly interacting particles, University of
  Chicago press, 1996.

\bibitem{1988PhRvD..37.1237R}
G.~{Raffelt}, L.~{Stodolsky}, {Mixing of the photon with low-mass particles},
  \prd 37~(5) (1988) 1237--1249.
\newblock \href {https://doi.org/10.1103/PhysRevD.37.1237}
  {\path{doi:10.1103/PhysRevD.37.1237}}.

\bibitem{2017NatPh..13..584A}
V.~{Anastassopoulos}, S.~{Aune}, K.~{Barth}, A.~{Belov}, H.~{Br{\"a}uninger},
  G.~{Cantatore}, J.~M. {Carmona}, J.~F. {Castel}, S.~A. {Cetin},
  F.~{Christensen}, J.~I. {Collar}, T.~{Dafni}, M.~{Davenport}, T.~A. {Decker},
  A.~{Dermenev}, K.~{Desch}, C.~{Eleftheriadis}, G.~{Fanourakis},
  E.~{Ferrer-Ribas}, H.~{Fischer}, J.~A. {Garc{\'\i}a}, A.~{Gardikiotis}, J.~G.
  {Garza}, E.~N. {Gazis}, T.~{Geralis}, I.~{Giomataris}, S.~{Gninenko}, C.~J.
  {Hailey}, M.~D. {Hasinoff}, D.~H.~H. {Hoffmann}, F.~J. {Iguaz}, I.~G.
  {Irastorza}, A.~{Jakobsen}, J.~{Jacoby}, K.~{Jakov{\v{c}}i{\'c}},
  J.~{Kaminski}, M.~{Karuza}, N.~{Kralj}, M.~{Kr{\v{c}}mar}, S.~{Kostoglou},
  C.~{Krieger}, B.~{Laki{\'c}}, J.~M. {Laurent}, A.~{Liolios},
  A.~{Ljubi{\v{c}}i{\'c}}, G.~{Luz{\'o}n}, M.~{Maroudas}, L.~{Miceli},
  S.~{Neff}, I.~{Ortega}, T.~{Papaevangelou}, K.~{Paraschou}, M.~J.
  {Pivovaroff}, G.~{Raffelt}, M.~{Rosu}, J.~{Ruz}, E.~R. {Ch{\'o}liz},
  I.~{Savvidis}, S.~{Schmidt}, Y.~K. {Semertzidis}, S.~K. {Solanki},
  L.~{Stewart}, T.~{Vafeiadis}, J.~K. {Vogel}, S.~C. {Yildiz}, K.~{Zioutas},
  {New CAST limit on the axion-photon interaction}, Nature Physics 13~(6)
  (2017) 584--590.
\newblock \href {http://arxiv.org/abs/1705.02290} {\path{arXiv:1705.02290}},
  \href {https://doi.org/10.1038/nphys4109} {\path{doi:10.1038/nphys4109}}.

\bibitem{2016PhRvL.116p1101A}
M.~{Ajello}, A.~{Albert}, B.~{Anderson}, L.~{Baldini}, G.~{Barbiellini},
  D.~{Bastieri}, R.~{Bellazzini}, E.~{Bissaldi}, R.~D. {Blandford}, E.~D.
  {Bloom}, R.~{Bonino}, E.~{Bottacini}, J.~{Bregeon}, P.~{Bruel}, R.~{Buehler},
  G.~A. {Caliandro}, R.~A. {Cameron}, M.~{Caragiulo}, P.~A. {Caraveo},
  C.~{Cecchi}, A.~{Chekhtman}, S.~{Ciprini}, J.~{Cohen-Tanugi}, J.~{Conrad},
  F.~{Costanza}, F.~{D'Ammando}, A.~{de Angelis}, F.~{de Palma}, R.~{Desiante},
  M.~{Di Mauro}, L.~{Di Venere}, A.~{Dom{\'\i}nguez}, P.~S. {Drell},
  C.~{Favuzzi}, W.~B. {Focke}, A.~{Franckowiak}, Y.~{Fukazawa}, S.~{Funk},
  P.~{Fusco}, F.~{Gargano}, D.~{Gasparrini}, N.~{Giglietto}, T.~{Glanzman},
  G.~{Godfrey}, S.~{Guiriec}, D.~{Horan}, G.~{J{\'o}hannesson},
  M.~{Katsuragawa}, S.~{Kensei}, M.~{Kuss}, S.~{Larsson}, L.~{Latronico},
  J.~{Li}, L.~{Li}, F.~{Longo}, F.~{Loparco}, P.~{Lubrano}, G.~M. {Madejski},
  S.~{Maldera}, A.~{Manfreda}, M.~{Mayer}, M.~N. {Mazziotta}, M.~{Meyer}, P.~F.
  {Michelson}, N.~{Mirabal}, T.~{Mizuno}, M.~E. {Monzani}, A.~{Morselli}, I.~V.
  {Moskalenko}, S.~{Murgia}, M.~{Negro}, E.~{Nuss}, C.~{Okada}, E.~{Orlando},
  J.~F. {Ormes}, D.~{Paneque}, J.~S. {Perkins}, M.~{Pesce-Rollins}, F.~{Piron},
  G.~{Pivato}, T.~A. {Porter}, S.~{Rain{\`o}}, R.~{Rando}, M.~{Razzano},
  A.~{Reimer}, M.~{S{\'a}nchez-Conde}, C.~{Sgr{\`o}}, D.~{Simone}, E.~J.
  {Siskind}, F.~{Spada}, G.~{Spandre}, P.~{Spinelli}, H.~{Takahashi}, J.~B.
  {Thayer}, D.~F. {Torres}, G.~{Tosti}, E.~{Troja}, Y.~{Uchiyama}, K.~S.
  {Wood}, M.~{Wood}, G.~{Zaharijas}, S.~{Zimmer}, {Fermi-LAT Collaboration},
  {Search for Spectral Irregularities due to Photon-Axionlike-Particle
  Oscillations with the Fermi Large Area Telescope}, \prl 116~(16) (2016)
  161101.
\newblock \href {http://arxiv.org/abs/1603.06978} {\path{arXiv:1603.06978}},
  \href {https://doi.org/10.1103/PhysRevLett.116.161101}
  {\path{doi:10.1103/PhysRevLett.116.161101}}.

\bibitem{2017ApJ...847..101B}
M.~{Berg}, J.~P. {Conlon}, F.~{Day}, N.~{Jennings}, S.~{Krippendorf}, A.~J.
  {Powell}, M.~{Rummel}, {Constraints on Axion-like Particles from X-Ray
  Observations of NGC1275}, \apj 847~(2) (2017) 101.
\newblock \href {http://arxiv.org/abs/1605.01043} {\path{arXiv:1605.01043}},
  \href {https://doi.org/10.3847/1538-4357/aa8b16}
  {\path{doi:10.3847/1538-4357/aa8b16}}.

\bibitem{2018MNRAS.479.2243C}
L.~{Chen}, J.~P. {Conlon}, {Constraints on massive axion-like particles from
  X-ray observations of NGC 1275}, \mnras 479~(2) (2018) 2243--2248.
\newblock \href {http://arxiv.org/abs/1712.08313} {\path{arXiv:1712.08313}},
  \href {https://doi.org/10.1093/mnras/sty1591}
  {\path{doi:10.1093/mnras/sty1591}}.

\bibitem{2020PhLB..80235252L}
M.~{Libanov}, S.~{Troitsky}, {On the impact of magnetic-field models in galaxy
  clusters on constraints on axion-like particles from the lack of
  irregularities in high-energy spectra of astrophysical sources}, Physics
  Letters B 802 (2020) 135252.
\newblock \href {http://arxiv.org/abs/1908.03084} {\path{arXiv:1908.03084}},
  \href {https://doi.org/10.1016/j.physletb.2020.135252}
  {\path{doi:10.1016/j.physletb.2020.135252}}.

\bibitem{2011PhRvD..84l5019F}
M.~{Fairbairn}, T.~{Rashba}, S.~{Troitsky}, {Photon-axion mixing and ultra-high
  energy cosmic rays from BL Lac type objects: Shining light through the
  Universe}, \prd 84~(12) (2011) 125019.
\newblock \href {http://arxiv.org/abs/0901.4085} {\path{arXiv:0901.4085}},
  \href {https://doi.org/10.1103/PhysRevD.84.125019}
  {\path{doi:10.1103/PhysRevD.84.125019}}.

\bibitem{2019MNRAS.487..123G}
G.~{Galanti}, F.~{Tavecchio}, M.~{Roncadelli}, C.~{Evoli}, {Blazar VHE spectral
  alterations induced by photon-ALP oscillations}, \mnras 487~(1) (2019)
  123--132.
\newblock \href {http://arxiv.org/abs/1811.03548} {\path{arXiv:1811.03548}},
  \href {https://doi.org/10.1093/mnras/stz1144}
  {\path{doi:10.1093/mnras/stz1144}}.

\bibitem{Payez:2014xsa}
A.~Payez, C.~Evoli, T.~Fischer, M.~Giannotti, A.~Mirizzi, A.~Ringwald,
  {Revisiting the SN1987A gamma-ray limit on ultralight axion-like particles},
  JCAP 02 (2015) 006.
\newblock \href {http://arxiv.org/abs/1410.3747} {\path{arXiv:1410.3747}},
  \href {https://doi.org/10.1088/1475-7516/2015/02/006}
  {\path{doi:10.1088/1475-7516/2015/02/006}}.

\bibitem{2011PhRvD..84h5001G}
R.~{Gill}, J.~S. {Heyl}, {Constraining the photon-axion coupling constant with
  magnetic white dwarfs}, \prd 84~(8) (2011) 085001.
\newblock \href {http://arxiv.org/abs/1105.2083} {\path{arXiv:1105.2083}},
  \href {https://doi.org/10.1103/PhysRevD.84.085001}
  {\path{doi:10.1103/PhysRevD.84.085001}}.

\bibitem{2009JETP..108..384P}
M.~S. {Pshirkov}, S.~B. {Popov}, {Conversion of dark matter axions to photons
  in magnetospheres of neutron stars}, Soviet Journal of Experimental and
  Theoretical Physics 108~(3) (2009) 384--388.
\newblock \href {http://arxiv.org/abs/0711.1264} {\path{arXiv:0711.1264}},
  \href {https://doi.org/10.1134/S1063776109030030}
  {\path{doi:10.1134/S1063776109030030}}.

\bibitem{2020PhRvL.125q1301F}
J.~W. {Foster}, Y.~{Kahn}, O.~{Macias}, Z.~{Sun}, R.~P. {Eatough}, V.~I.
  {Kondratiev}, W.~M. {Peters}, C.~{Weniger}, B.~R. {Safdi}, {Green Bank and
  Effelsberg Radio Telescope Searches for Axion Dark Matter Conversion in
  Neutron Star Magnetospheres}, \prl 125~(17) (2020) 171301.
\newblock \href {http://arxiv.org/abs/2004.00011} {\path{arXiv:2004.00011}},
  \href {https://doi.org/10.1103/PhysRevLett.125.171301}
  {\path{doi:10.1103/PhysRevLett.125.171301}}.

\bibitem{RAFFELT19901}
G.~G. Raffelt,
  \href{https://www.sciencedirect.com/science/article/pii/0370157390900546}{Astrophysical
  methods to constrain axions and other novel particle phenomena}, Physics
  Reports 198~(1) (1990) 1--113.
\newblock \href {https://doi.org/https://doi.org/10.1016/0370-1573(90)90054-6}
  {\path{doi:https://doi.org/10.1016/0370-1573(90)90054-6}}.
\newline\urlprefix\url{https://www.sciencedirect.com/science/article/pii/0370157390900546}

\bibitem{2019PhRvD..99d3011S}
A.~{Sedrakian}, {Axion cooling of neutron stars. II. Beyond hadronic axions},
  \prd 99~(4) (2019) 043011.
\newblock \href {http://arxiv.org/abs/1810.00190} {\path{arXiv:1810.00190}},
  \href {https://doi.org/10.1103/PhysRevD.99.043011}
  {\path{doi:10.1103/PhysRevD.99.043011}}.

\bibitem{2018JHEP...06..048F}
J.-F. {Fortin}, K.~{Sinha}, {Constraining axion-like-particles with hard X-ray
  emission from magnetars}, Journal of High Energy Physics 2018~(6) (2018) 48.
\newblock \href {http://arxiv.org/abs/1804.01992} {\path{arXiv:1804.01992}},
  \href {https://doi.org/10.1007/JHEP06(2018)048}
  {\path{doi:10.1007/JHEP06(2018)048}}.

\bibitem{2019JHEP...01..163F}
J.-F. {Fortin}, K.~{Sinha}, {X-ray polarization signals from magnetars with
  axion-like-particles}, Journal of High Energy Physics 2019~(1) (2019) 163.
\newblock \href {http://arxiv.org/abs/1807.10773} {\path{arXiv:1807.10773}},
  \href {https://doi.org/10.1007/JHEP01(2019)163}
  {\path{doi:10.1007/JHEP01(2019)163}}.

\bibitem{2021PhRvD.103b3010L}
S.~J. {Lloyd}, P.~M. {Chadwick}, A.~M. {Brown}, H.-K. {Guo}, K.~{Sinha}, {Axion
  constraints from quiescent soft gamma-ray emission from magnetars}, \prd
  103~(2) (2021) 023010.
\newblock \href {http://arxiv.org/abs/2001.10849} {\path{arXiv:2001.10849}},
  \href {https://doi.org/10.1103/PhysRevD.103.023010}
  {\path{doi:10.1103/PhysRevD.103.023010}}.

\bibitem{2006PhRvD..74l3003L}
D.~{Lai}, J.~{Heyl}, {Probing axions with radiation from magnetic stars}, \prd
  74~(12) (2006) 123003.
\newblock \href {http://arxiv.org/abs/astro-ph/0609775}
  {\path{arXiv:astro-ph/0609775}}, \href
  {https://doi.org/10.1103/PhysRevD.74.123003}
  {\path{doi:10.1103/PhysRevD.74.123003}}.

\bibitem{2012ApJ...748..116P}
R.~{Perna}, W.~C.~G. {Ho}, L.~{Verde}, M.~{van Adelsberg}, R.~{Jimenez},
  {Signatures of Photon-Axion Conversion in the Thermal Spectra and
  Polarization of Neutron Stars}, \apj 748~(2) (2012) 116.
\newblock \href {http://arxiv.org/abs/1201.5390} {\path{arXiv:1201.5390}},
  \href {https://doi.org/10.1088/0004-637X/748/2/116}
  {\path{doi:10.1088/0004-637X/748/2/116}}.

\bibitem{2009ApJ...705..798K}
D.~L. {Kaplan}, M.~H. {van Kerkwijk}, {Constraining the Spin-down of the Nearby
  Isolated Neutron Star RX J0806.4-4123, and Implications for the Population of
  Nearby Neutron Stars}, \apj 705~(1) (2009) 798--808.
\newblock \href {http://arxiv.org/abs/0909.5218} {\path{arXiv:0909.5218}},
  \href {https://doi.org/10.1088/0004-637X/705/1/798}
  {\path{doi:10.1088/0004-637X/705/1/798}}.

\bibitem{2009ASSL..357..141T}
R.~Turolla, \href{https://doi.org/10.1007/978-3-540-76965-1_7}{Isolated Neutron
  Stars: The Challenge of Simplicity}, Springer Berlin Heidelberg, Berlin,
  Heidelberg, 2009, Ch.~7, pp. 141--163.
\newblock \href {https://doi.org/10.1007/978-3-540-76965-1_7}
  {\path{doi:10.1007/978-3-540-76965-1_7}}.
\newline\urlprefix\url{https://doi.org/10.1007/978-3-540-76965-1_7}

\bibitem{2014PhyU...57..735P}
A.~Y. {Potekhin}, {Atmospheres and radiating surfaces of neutron stars},
  Physics Uspekhi 57~(8) (2014) 735--770.
\newblock \href {http://arxiv.org/abs/1403.0074} {\path{arXiv:1403.0074}},
  \href {https://doi.org/10.3367/UFNe.0184.201408a.0793}
  {\path{doi:10.3367/UFNe.0184.201408a.0793}}.

\bibitem{2016MNRAS.459.3585G}
D.~{Gonz{\'a}lez Caniulef}, S.~{Zane}, R.~{Taverna}, R.~{Turolla}, K.~{Wu},
  {Polarized thermal emission from X-ray dim isolated neutron stars: the case
  of RX J1856.5-3754}, \mnras 459~(4) (2016) 3585--3595.
\newblock \href {http://arxiv.org/abs/1604.01552} {\path{arXiv:1604.01552}},
  \href {https://doi.org/10.1093/mnras/stw804}
  {\path{doi:10.1093/mnras/stw804}}.

\bibitem{2006RPPh...69.2631H}
A.~K. {Harding}, D.~{Lai}, {Physics of strongly magnetized neutron stars},
  Reports on Progress in Physics 69~(9) (2006) 2631--2708.
\newblock \href {http://arxiv.org/abs/astro-ph/0606674}
  {\path{arXiv:astro-ph/0606674}}, \href
  {https://doi.org/10.1088/0034-4885/69/9/R03}
  {\path{doi:10.1088/0034-4885/69/9/R03}}.

\bibitem{2006MNRAS.373.1495V}
M.~{van Adelsberg}, D.~{Lai}, {Atmosphere models of magnetized neutron stars:
  QED effects, radiation spectra and polarization signals}, \mnras 373~(4)
  (2006) 1495--1522.
\newblock \href {http://arxiv.org/abs/astro-ph/0607168}
  {\path{arXiv:astro-ph/0607168}}, \href
  {https://doi.org/10.1111/j.1365-2966.2006.11098.x}
  {\path{doi:10.1111/j.1365-2966.2006.11098.x}}.

\bibitem{1997ApJ...491..270L}
D.~{Lai}, E.~E. {Salpeter}, {Hydrogen Phases on the Surfaces of a Strongly
  Magnetized Neutron Star}, \apj 491~(1) (1997) 270--285.
\newblock \href {http://arxiv.org/abs/astro-ph/9704130}
  {\path{arXiv:astro-ph/9704130}}, \href {https://doi.org/10.1086/304937}
  {\path{doi:10.1086/304937}}.

\bibitem{2001RvMP...73..629L}
D.~{Lai}, {Matter in strong magnetic fields}, Reviews of Modern Physics 73~(3)
  (2001) 629.
\newblock \href {http://arxiv.org/abs/astro-ph/0009333}
  {\path{arXiv:astro-ph/0009333}}, \href
  {https://doi.org/10.1103/RevModPhys.73.629}
  {\path{doi:10.1103/RevModPhys.73.629}}.

\bibitem{2007MNRAS.382.1833M}
Z.~{Medin}, D.~{Lai}, {Condensed surfaces of magnetic neutron stars, thermal
  surface emission, and particle acceleration above pulsar polar caps}, \mnras
  382~(4) (2007) 1833--1852.
\newblock \href {https://doi.org/10.1111/j.1365-2966.2007.12492.x}
  {\path{doi:10.1111/j.1365-2966.2007.12492.x}}.

\bibitem{2010A&A...522A.111S}
V.~{Suleimanov}, V.~{Hambaryan}, A.~Y. {Potekhin}, M.~{van Adelsberg},
  R.~{Neuh{\"a}user}, K.~{Werner}, {Radiative properties of highly magnetized
  isolated neutron star surfaces and approximate treatment of absorption
  features in their spectra}, \aap 522 (2010) A111.
\newblock \href {http://arxiv.org/abs/1006.3292} {\path{arXiv:1006.3292}},
  \href {https://doi.org/10.1051/0004-6361/200913641}
  {\path{doi:10.1051/0004-6361/200913641}}.

\bibitem{2002PhRvD..66b3002H}
J.~S. {Heyl}, N.~J. {Shaviv}, {QED and the high polarization of the thermal
  radiation from neutron stars}, \prd 66~(2) (2002) 023002.
\newblock \href {http://arxiv.org/abs/astro-ph/0203058}
  {\path{arXiv:astro-ph/0203058}}, \href
  {https://doi.org/10.1103/PhysRevD.66.023002}
  {\path{doi:10.1103/PhysRevD.66.023002}}.

\bibitem{2003MNRAS.342..134H}
J.~S. {Heyl}, N.~J. {Shaviv}, D.~{Lloyd}, {The high-energy
  polarization-limiting radius of neutron star magnetospheres - I. Slowly
  rotating neutron stars}, \mnras 342~(1) (2003) 134--144.
\newblock \href {http://arxiv.org/abs/astro-ph/0302118}
  {\path{arXiv:astro-ph/0302118}}, \href
  {https://doi.org/10.1046/j.1365-8711.2003.06521.x}
  {\path{doi:10.1046/j.1365-8711.2003.06521.x}}.

\bibitem{2015MNRAS.454.3254T}
R.~{Taverna}, R.~{Turolla}, D.~{Gonzalez Caniulef}, S.~{Zane}, F.~{Muleri},
  P.~{Soffitta}, {Polarization of neutron star surface emission: a systematic
  analysis}, \mnras 454~(3) (2015) 3254--3266.
\newblock \href {http://arxiv.org/abs/1509.05023} {\path{arXiv:1509.05023}},
  \href {https://doi.org/10.1093/mnras/stv2168}
  {\path{doi:10.1093/mnras/stv2168}}.

\bibitem{2002ApJ...566L..85B}
A.~M. {Beloborodov}, {Gravitational Bending of Light Near Compact Objects},
  \apjl 566~(2) (2002) L85--L88.
\newblock \href {http://arxiv.org/abs/astro-ph/0201117}
  {\path{arXiv:astro-ph/0201117}}, \href {https://doi.org/10.1086/339511}
  {\path{doi:10.1086/339511}}.

\bibitem{2014MNRAS.438.1686T}
R.~{Taverna}, F.~{Muleri}, R.~{Turolla}, P.~{Soffitta}, S.~{Fabiani},
  L.~{Nobili}, {Probing magnetar magnetosphere through X-ray polarization
  measurements}, \mnras 438~(2) (2014) 1686--1697.
\newblock \href {http://arxiv.org/abs/1311.7500} {\path{arXiv:1311.7500}},
  \href {https://doi.org/10.1093/mnras/stt2310}
  {\path{doi:10.1093/mnras/stt2310}}.

\bibitem{1971AnPhy..67..599A}
S.~L. {Adler}, {Photon splitting and photon dispersion in a strong magnetic
  field.}, Annals of Physics 67 (1971) 599--647.
\newblock \href {https://doi.org/10.1016/0003-4916(71)90154-0}
  {\path{doi:10.1016/0003-4916(71)90154-0}}.

\bibitem{2004ApJ...612.1034P}
A.~Y. {Potekhin}, D.~{Lai}, G.~{Chabrier}, W.~C.~G. {Ho}, {Electromagnetic
  Polarization in Partially Ionized Plasmas with Strong Magnetic Fields and
  Neutron Star Atmosphere Models}, \apj 612~(2) (2004) 1034--1043.
\newblock \href {http://arxiv.org/abs/astro-ph/0405383}
  {\path{arXiv:astro-ph/0405383}}, \href {https://doi.org/10.1086/422679}
  {\path{doi:10.1086/422679}}.

\bibitem{1969ApJ...157..869G}
P.~{Goldreich}, W.~H. {Julian}, {Pulsar Electrodynamics}, \apj 157 (1969) 869.
\newblock \href {https://doi.org/10.1086/150119} {\path{doi:10.1086/150119}}.

\bibitem{2006MNRAS.368.1377S}
R.~M. {Shannon}, J.~S. {Heyl}, {Magnetospheric birefringence induces
  polarization signatures in neutron-star spectra}, \mnras 368~(3) (2006)
  1377--1380.
\newblock \href {http://arxiv.org/abs/astro-ph/0410548}
  {\path{arXiv:astro-ph/0410548}}, \href
  {https://doi.org/10.1111/j.1365-2966.2006.10208.x}
  {\path{doi:10.1111/j.1365-2966.2006.10208.x}}.

\bibitem{2007PhRvD..76l3011H}
K.~A. {Hochmuth}, G.~{Sigl}, {Effects of axion-photon mixing on gamma-ray
  spectra from magnetized astrophysical sources}, \prd 76~(12) (2007) 123011.
\newblock \href {http://arxiv.org/abs/0708.1144} {\path{arXiv:0708.1144}},
  \href {https://doi.org/10.1103/PhysRevD.76.123011}
  {\path{doi:10.1103/PhysRevD.76.123011}}.

\bibitem{1986SvA....30..567M}
A.~G. {Muslimov}, A.~I. {Tsygan}, {Electric Fields Generated by a Rotating
  Neutron Star in a Vacuum with Allowance for Gtr Effects}, \sovast 30 (1986)
  567.

\bibitem{1996ApJ...473.1067P}
D.~{Page}, A.~{Sarmiento}, {Surface Temperature of a Magnetized Neutron Star
  and Interpretation of the ROSAT Data. II.}, \apj 473 (1996) 1067.
\newblock \href {http://arxiv.org/abs/astro-ph/9602042}
  {\path{arXiv:astro-ph/9602042}}, \href {https://doi.org/10.1086/178216}
  {\path{doi:10.1086/178216}}.

\bibitem{2011ApJ...736..117K}
D.~L. {Kaplan}, A.~{Kamble}, M.~H. {van Kerkwijk}, W.~C.~G. {Ho}, {New
  Optical/Ultraviolet Counterparts and the Spectral Energy Distributions of
  Nearby, Thermally Emitting, Isolated Neutron Stars}, \apj 736~(2) (2011) 117.
\newblock \href {http://arxiv.org/abs/1105.4178} {\path{arXiv:1105.4178}},
  \href {https://doi.org/10.1088/0004-637X/736/2/117}
  {\path{doi:10.1088/0004-637X/736/2/117}}.

\bibitem{2007MNRAS.377..905M}
K.~{Mori}, W.~C.~G. {Ho}, {Modelling mid-Z element atmospheres for strongly
  magnetized neutron stars}, \mnras 377~(2) (2007) 905--919.
\newblock \href {http://arxiv.org/abs/astro-ph/0611145}
  {\path{arXiv:astro-ph/0611145}}, \href
  {https://doi.org/10.1111/j.1365-2966.2007.11663.x}
  {\path{doi:10.1111/j.1365-2966.2007.11663.x}}.

\bibitem{2006A&A...459..175P}
J.~F. {P{\'e}rez-Azor{\'\i}n}, J.~A. {Pons}, J.~A. {Miralles}, G.~{Miniutti},
  {A self-consistent model of isolated neutron stars: the case of the X-ray
  pulsar RX J0720.4-3125}, \aap 459~(1) (2006) 175--185.
\newblock \href {http://arxiv.org/abs/astro-ph/0603752}
  {\path{arXiv:astro-ph/0603752}}, \href
  {https://doi.org/10.1051/0004-6361:20065827}
  {\path{doi:10.1051/0004-6361:20065827}}.

\bibitem{2007MNRAS.375..821H}
W.~C.~G. {Ho}, D.~L. {Kaplan}, P.~{Chang}, M.~{van Adelsberg}, A.~Y.
  {Potekhin}, {Magnetic hydrogen atmosphere models and the neutron star RX
  J1856.5-3754}, \mnras 375~(3) (2007) 821--830.
\newblock \href {http://arxiv.org/abs/astro-ph/0612145}
  {\path{arXiv:astro-ph/0612145}}, \href
  {https://doi.org/10.1111/j.1365-2966.2006.11376.x}
  {\path{doi:10.1111/j.1365-2966.2006.11376.x}}.

\bibitem{2011A&A...534A..74H}
V.~{Hambaryan}, V.~{Suleimanov}, A.~D. {Schwope}, R.~{Neuh{\"a}user},
  K.~{Werner}, A.~Y. {Potekhin}, {Phase-resolved spectroscopic study of the
  isolated neutron star RBS 1223 (1RXS J130848.6+212708)}, \aap 534 (2011) A74.
\newblock \href {https://doi.org/10.1051/0004-6361/201117548}
  {\path{doi:10.1051/0004-6361/201117548}}.

\bibitem{2017MNRAS.465..492M}
R.~P. {Mignani}, V.~{Testa}, D.~{Gonz{\'a}lez Caniulef}, R.~{Taverna},
  R.~{Turolla}, S.~{Zane}, K.~{Wu}, {Evidence for vacuum birefringence from the
  first optical-polarimetry measurement of the isolated neutron star RX
  J1856.5-3754}, \mnras 465~(1) (2017) 492--500.
\newblock \href {http://arxiv.org/abs/1610.08323} {\path{arXiv:1610.08323}},
  \href {https://doi.org/10.1093/mnras/stw2798}
  {\path{doi:10.1093/mnras/stw2798}}.

\bibitem{2002ApJ...574..332T}
C.~{Thompson}, M.~{Lyutikov}, S.~R. {Kulkarni}, {Electrodynamics of Magnetars:
  Implications for the Persistent X-Ray Emission and Spin-down of the Soft
  Gamma Repeaters and Anomalous X-Ray Pulsars}, \apj 574~(1) (2002) 332--355.
\newblock \href {http://arxiv.org/abs/astro-ph/0110677}
  {\path{arXiv:astro-ph/0110677}}, \href {https://doi.org/10.1086/340586}
  {\path{doi:10.1086/340586}}.

\bibitem{Beloborodov_2007}
A.~M. Beloborodov, C.~Thompson, \href{https://doi.org/10.1086/508917}{Corona of
  magnetars}, \apj 657~(2) (2007) 967--993.
\newblock \href {https://doi.org/10.1086/508917} {\path{doi:10.1086/508917}}.
\newline\urlprefix\url{https://doi.org/10.1086/508917}

\bibitem{2020ApJ...904..184T}
C.~{Thompson}, A.~{Kostenko}, {Pair Plasma in Super-QED Magnetic Fields and the
  Hard X-Ray/Optical Emission of Magnetars}, \apj 904~(2) (2020) 184.
\newblock \href {http://arxiv.org/abs/2008.08659} {\path{arXiv:2008.08659}},
  \href {https://doi.org/10.3847/1538-4357/abbe87}
  {\path{doi:10.3847/1538-4357/abbe87}}.

\bibitem{2015RPPh...78k6901T}
R.~{Turolla}, S.~{Zane}, A.~L. {Watts}, {Magnetars: the physics behind
  observations. A review}, Reports on Progress in Physics 78~(11) (2015)
  116901.
\newblock \href {http://arxiv.org/abs/1507.02924} {\path{arXiv:1507.02924}},
  \href {https://doi.org/10.1088/0034-4885/78/11/116901}
  {\path{doi:10.1088/0034-4885/78/11/116901}}.

\end{thebibliography}

\end{document}